\documentclass[12pt]{article}
\usepackage[dvips]{graphicx}
\usepackage{epsfig}
\usepackage{amsmath,amssymb,amsbsy}
\usepackage{amsfonts}
\begin{document}
\thispagestyle{empty}
\begin{center}

{\Large\bf{
New developments in the statistical approach of
parton distributions \footnote{Presented by J. Soffer at EDS Blois 2015: The 16th conference on Elastic 
and Diffractive scattering, 06/29/15 - 04/07/15, Borgo, Corsica, France}}}
\vskip1.4cm
{\bf Claude Bourrely}
\vskip 0.3cm
Aix-Marseille Universit\'e, D\'epartement de Physique, \\
Facult\'e des Sciences site de Luminy, 13288 Marseille, Cedex 09, France\\
\vskip 0.5cm
{\bf Jacques Soffer}
\vskip 0.3cm
Physics Department, Temple University,\\
1925 N, 12th Street, Philadelphia, PA 19122-1801, USA
\vskip 1.5cm
{\bf Abstract}\end{center}

The quantum statistical parton distributions approach proposed more than one decade
ago is revisited by considering a larger set of recent and accurate Deep Inelastic 
Scattering experimental results. It enables us to improve the description of the data 
by means of a new determination of the parton distributions. This global next-to-leading 
order QCD analysis leads to a good description of several structure functions, involving 
unpolarized parton distributions and helicity distributions, in terms of a rather small 
number of free parameters. There are several challenging issues. The predictions 
of this theoretical approach will be tested for single-jet production and charge asymmetry 
in $W^{\pm}$ production in $\bar p p$ and $p p$ collisions up to LHC energies, using recent 
data and also for forthcoming experimental results.\\\\

  
\section{Introduction}
Deep Inelastic Scattering (DIS) of leptons and nucleons is indeed our main
source of information to study the internal nucleon structure in terms of
parton distributions. Several years ago a new set of parton distribution
functions (PDFs) was constructed in the framework of a statistical approach 
of the nucleon \cite{bbs1}. For quarks (antiquarks), the building blocks 
are the helicity dependent distributions $q^{\pm}(x)$ ($\bar q^{\pm}(x)$). 
This allows to describe simultaneously the unpolarized distributions 
$q(x)= q^{+}(x)+q^{-}(x)$ and the helicity distributions $\Delta q(x) = q^{+}(x)-q^{-}(x)$ 
(similarly for antiquarks). At the initial energy scale $Q_0^2$, these distributions 
are given by the sum of two terms, a quasi Fermi-Dirac function and a helicity 
independent diffractive contribution. The flavor asymmetry for the light sea, 
{\it i.e.} $\bar d (x) >\bar u (x)$, observed in the data is built in. This is simply 
understood in terms of the Pauli exclusion principle, based on the fact that the proton contains
two up-quarks and only one down-quark. We predict that $\bar d(x) /\bar u(x)$ 
must remain above one for all $x$ values and this is a real challenge for our approach, 
in particular in the large $x$ region which is under experimental investigation at the moment. 
The flattening out of the ratio $d(x)/u(x)$ in the high $x$ region, predicted by the 
statistical approach, is another interesting challenge worth mentioning. The chiral 
properties of QCD lead to strong relations between $q(x)$ and $\bar q (x)$.
For example, it is found that the well established result $\Delta u (x)>0 $\
implies $\Delta \bar u (x)>0$ and similarly $\Delta d (x)<0$ leads to $\Delta \bar d (x)<0$. 
This earlier prediction was confirmed by recent polarized DIS data and it was also 
demonstrated that the magnitude predicted by the statistical approach is compatible 
with recent BNL-RHIC data on $W^{\pm}$ production \cite{bbsW}. In addition we found the
approximate equality of the flavor asymmetries, namely $\bar d(x) - \bar u(x)
\sim \Delta \bar u(x) - \Delta \bar d(x)$. Concerning the gluon, the unpolarized 
distribution $G(x,Q_0^2)$ is given in terms of a quasi Bose-Einstein function, 
with only {\it one free parameter}. The new analysis of a larger set of recent 
accurate DIS data leads to the emergence of a large positive gluon helicity
distribution, giving a significant contribution to the proton spin, a major
point which was emphasized in a recent letter \cite{bs14}.\\ 
It is crucial to note that the quantum-statistical approach differs from the usual 
global parton fitting methodology for the following reasons:\\
i) It incorporates physical principles to reduce the number of free parameters which 
have a physical interpretation\\
ii) It has very specific predictions, so far confirmed by the data\\
iii) It is an attempt to reach a more physical picture on our knowledge of the nucleon 
structure, the ultimate goal being to solve the problem of confinement\\
iv) Treating simultaneously unpolarized distributions and helicity distributions, a unique 
siuation in the literature, has the advantage to give access to a vast set of experimental data, 
in particular up to LHC energies\\

\section{Review of the statistical parton distributions}
Let us now recall the main features of the statistical approach for building up
the PDFs, as opposed to the standard polynomial type
parameterizations of the PDF, based on Regge theory at low $x$ and on counting
rules at large $x$.
The fermion distributions are given by the sum of two terms,
a quasi Fermi-Dirac function and a helicity independent diffractive
contribution, at the input energy scale $Q_0^2=1 \mbox{GeV}^2$,
\begin{equation}
xq^h(x,Q^2_0)=
\frac{A_{q}X^h_{0q}x^{b_q}}{\exp [(x-X^h_{0q})/\bar{x}]+1}+
\frac{\tilde{A}_{q}x^{\tilde{b}_{q}}}{\exp(x/\bar{x})+1}~,
\label{eq1}
\end{equation}
\begin{equation}
x\bar{q}^h(x,Q^2_0)=
\frac{{\bar A_{q}}(X^{-h}_{0q})^{-1}x^{\bar{b}_ q}}{\exp
[(x+X^{-h}_{0q})/\bar{x}]+1}+
\frac{\tilde{A}_{q}x^{\tilde{b}_{q}}}{\exp(x/\bar{x})+1}~.
\label{eq2}
\end{equation}
We note that the diffractive term is absent in the quark helicity distribution $\Delta q$ 
and in the quark valence contribution $q - \bar q$.\\
In Eqs.~(\ref{eq1},\ref{eq2}) the multiplicative factors $X^{h}_{0q}$ and
$(X^{-h}_{0q})^{-1}$ in
the numerators of the non-diffractive parts of the $q$'s and $\bar{q}$'s
distributions, imply a modification
of the quantum statistical form, we were led to propose in order to agree with
experimental data. The presence of these multiplicative factors was justified
in our earlier attempt to generate the transverse momentum dependence (TMD)
\cite{bbs5, bbs6}.
The parameter $\bar{x}$ plays the role of a {\it universal temperature}
and $X^{\pm}_{0q}$ are the two {\it thermodynamical potentials} of the quark
$q$, with helicity $h=\pm$. They represent the fundamental characteristics of
the model. Notice the change of sign of the potentials
and helicity for the antiquarks \footnote{~At variance with statistical
mechanics where the distributions are expressed in terms of the energy, here
one uses
 $x$ which is clearly the natural variable entering in all the sum rules of the
parton model.}.\\
For a given flavor $q$ the corresponding quark and antiquark distributions
involve {\it eight} free parameters: $X^{\pm}_{0q}$, $A_q$, $\bar {A}_q$,
$\tilde {A}_q$, $b_q$, $\bar {b}_q$ and $\tilde {b}_q$. It reduces to $\it
seven$ since one of them is fixed by the valence sum rule, $\int (q(x) - \bar
{q}(x))dx = N_q$, where $N_q = 2, 1, 0 ~~\mbox{for}~~ u, d, s$, respectively.

For the light quarks $q=u,d$,  the total number of free parameters is reduced
to $\it eight$ by taking, as in Ref. \cite{bbs1}, $A_u=A_d$, $\bar {A}_u = \bar
{A}_d$, $\tilde {A}_u = \tilde {A}_d$, $b_u = b_d$, $\bar {b}_u = \bar {b}_d$
and $\tilde {b}_u = \tilde {b}_d$. For the strange quark and antiquark
distributions, the simple choice made in Ref. \cite{bbs1}
was improved in Ref. \cite{bbs2}, but here they are expressed in terms of $\it
seven$ free parameters.\\
For the gluons we consider the black-body inspired expression
\begin{equation}
xG(x,Q^2_0) = \frac{A_Gx^{b_G}}{\exp(x/\bar{x})-1}~,
\label{eq3}
\end{equation}
a quasi Bose-Einstein function, with $b_G$ being the only free parameter, since
$A_G$ is determined by the momentum sum rule.
In our earlier works \cite{bbs1,bbs4}, we were assuming that, at the input
energy scale, the helicity gluon distribution vanishes, so
\begin{equation}
x\Delta G(x,Q^2_0)=0~.
\label{eq4}
\end{equation}
However as a result of the present analysis of a much larger set of very
accurate unpolarized and polarized DIS data, we must give up this simplifying
assumption. We are now taking

\begin{equation}
 x\Delta G(x,Q^2_0) = \frac {\tilde A_G x^{\tilde b_G}}{(1+ c_G
x^{d_G})}\!\cdot\!\frac{1}{\exp(x/\bar x - 1) } \,.
 \end{equation}

To summarize the new determination of all PDFs involves a total of {\it twenty
one} free parameters: in addition to the temperature $\bar x$ and the exponent
$b_G$ of the gluon distribution, we have {\it eight} free parameters for the
light quarks $(u,d)$, {\it seven} free parameters for the strange quarks and
{\it four} free parameters for the gluon helicity distribution. These
parameters have been determined from a next-to-leading order (NLO) QCD fit of a
large set of accurate DIS data,  unpolarized and polarized structure functions \cite{bs15}.

\section{A selection of results}
\begin{figure}[hbp]  
\begin{center}
\includegraphics[width=5.8cm]{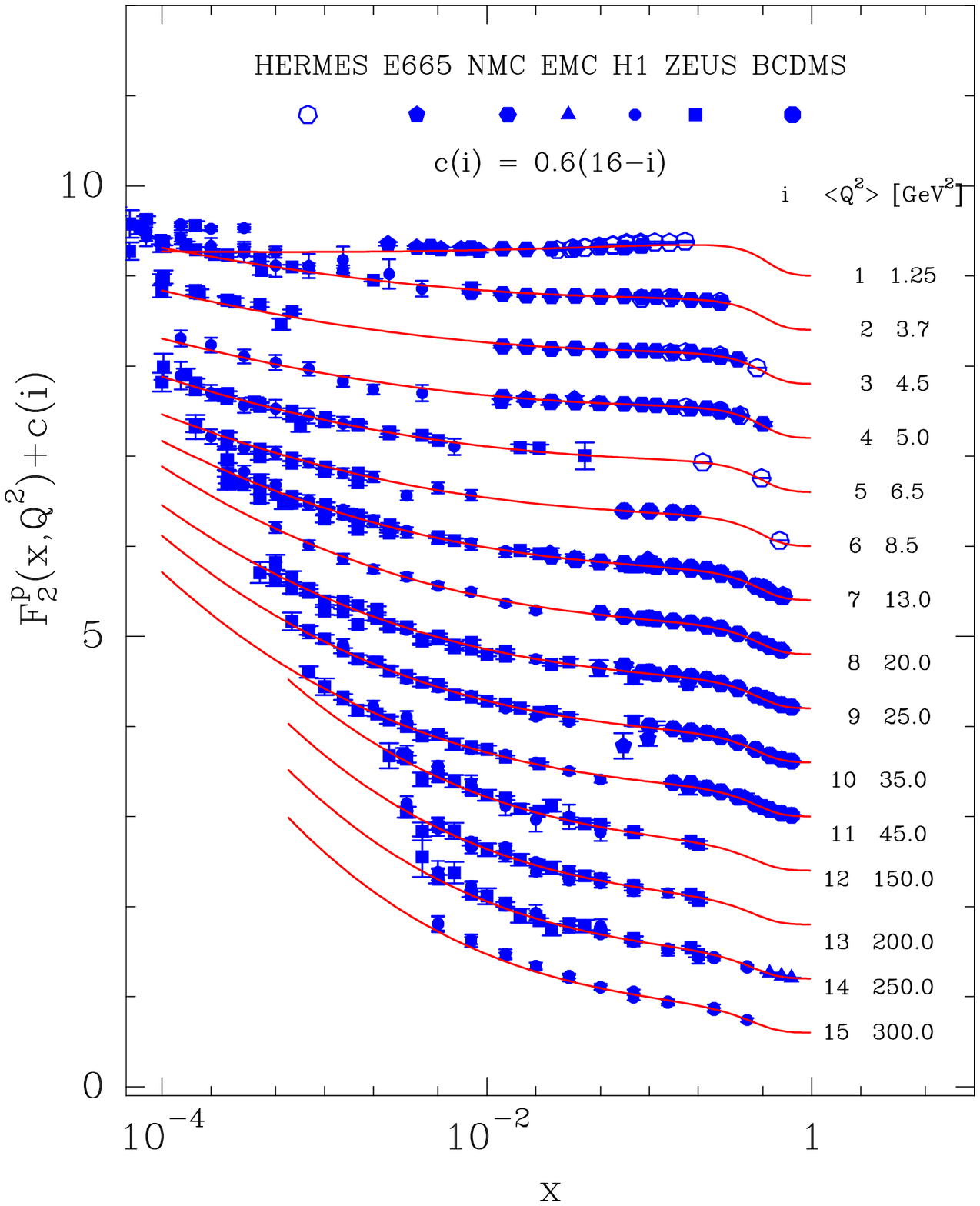}
\includegraphics[width=5.8cm]{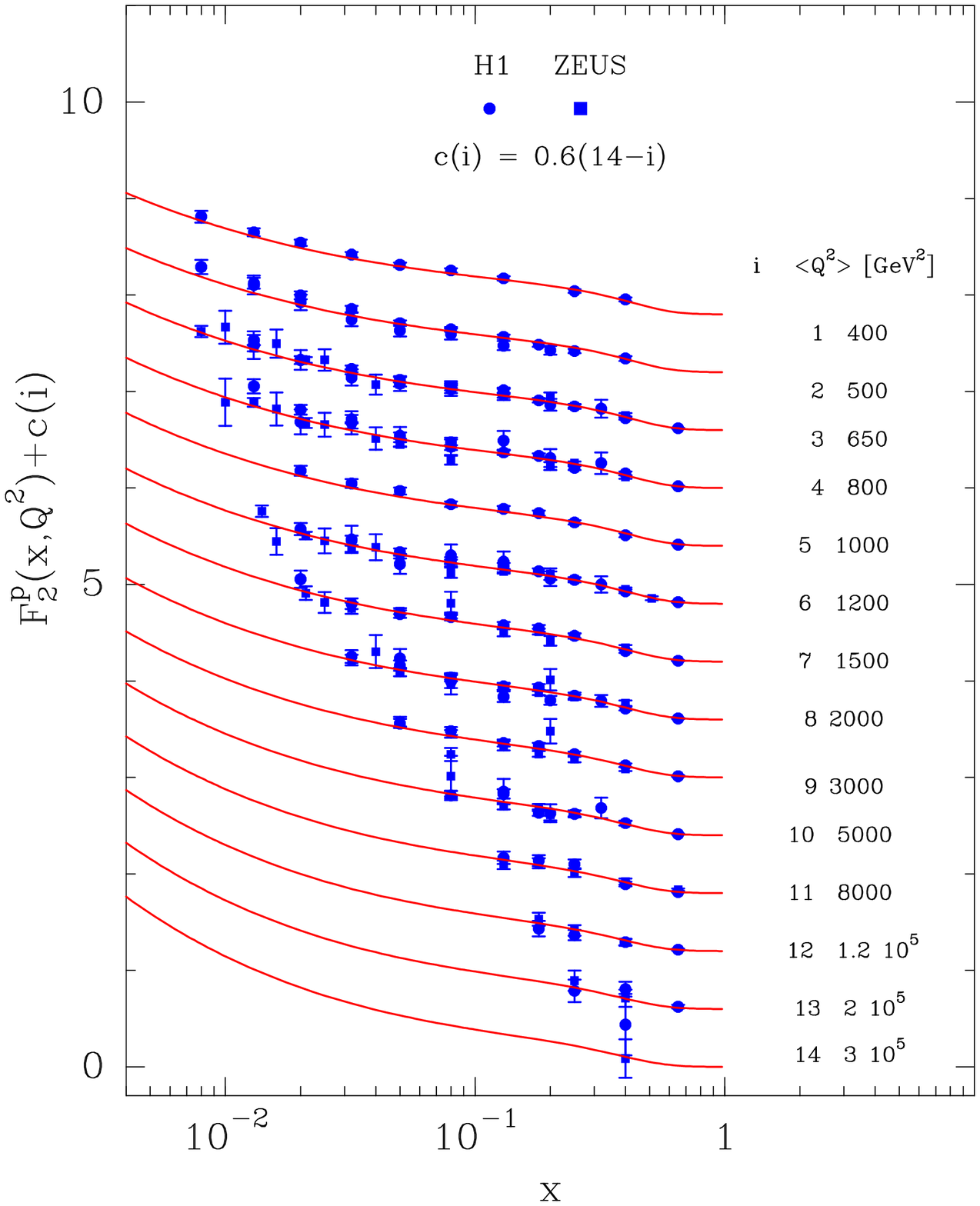}
\caption[*]{\baselineskip 1pt
 $F^{p}_{2}(x,Q^2)$ as a function of $x$ for fixed $\langle Q^2 \rangle$ and data from 
HERMES, E665, NMC, EMC, H1, ZEUS, BCDMS. {\it Left}: The function $c(i) = 0.6(16 - i )$, 
$i=1$ corresponds to $\langle Q^2 \rangle = 1.25 \mbox{GeV}^2$. {\it Right}: The function 
$c(i) = 0.6(14 - i )$, $i=1$ corresponds to $\langle Q^2 \rangle = 400\mbox{GeV}^2$. 
The curves are the results of the statistical approach.}
\label{f2p}
\end{center}
\end{figure}

Some selected experimental tests for the unpolarized PDFs have been considered from
$\mu N$ and $e N$ DIS, for which several experiments have yielded a
large number of data points on the structure functions $F_2^N(x,Q^2)$, $N$
stands for either a proton or a deuterium target. We have used fixed target
measurements which probe a rather limited kinematic region in $Q^2$ and $x$ and
also HERA data which cover a very large $Q^2$ range and probe the very low $x$
region, dominated by a fast rising  behavior, consistent with our diffractive
term (See Eq. (\ref {eq1})).\\
For illustration of the quality of our fit and, as an example, we show in
Fig.~\ref{f2p}, our results for $F_2 ^p (x,Q^2)$ on
different fixed proton targets, together with H1 and ZEUS data. We note that
the analysis of the scaling violations leads to a gluon distribution
$xG(x,Q^2)$, in fairly good agreement with our simple parameterization (See Eq.
(\ref{eq3})). 

\begin{figure}[hbp]  
\begin{center}
\includegraphics[width=6.5cm]{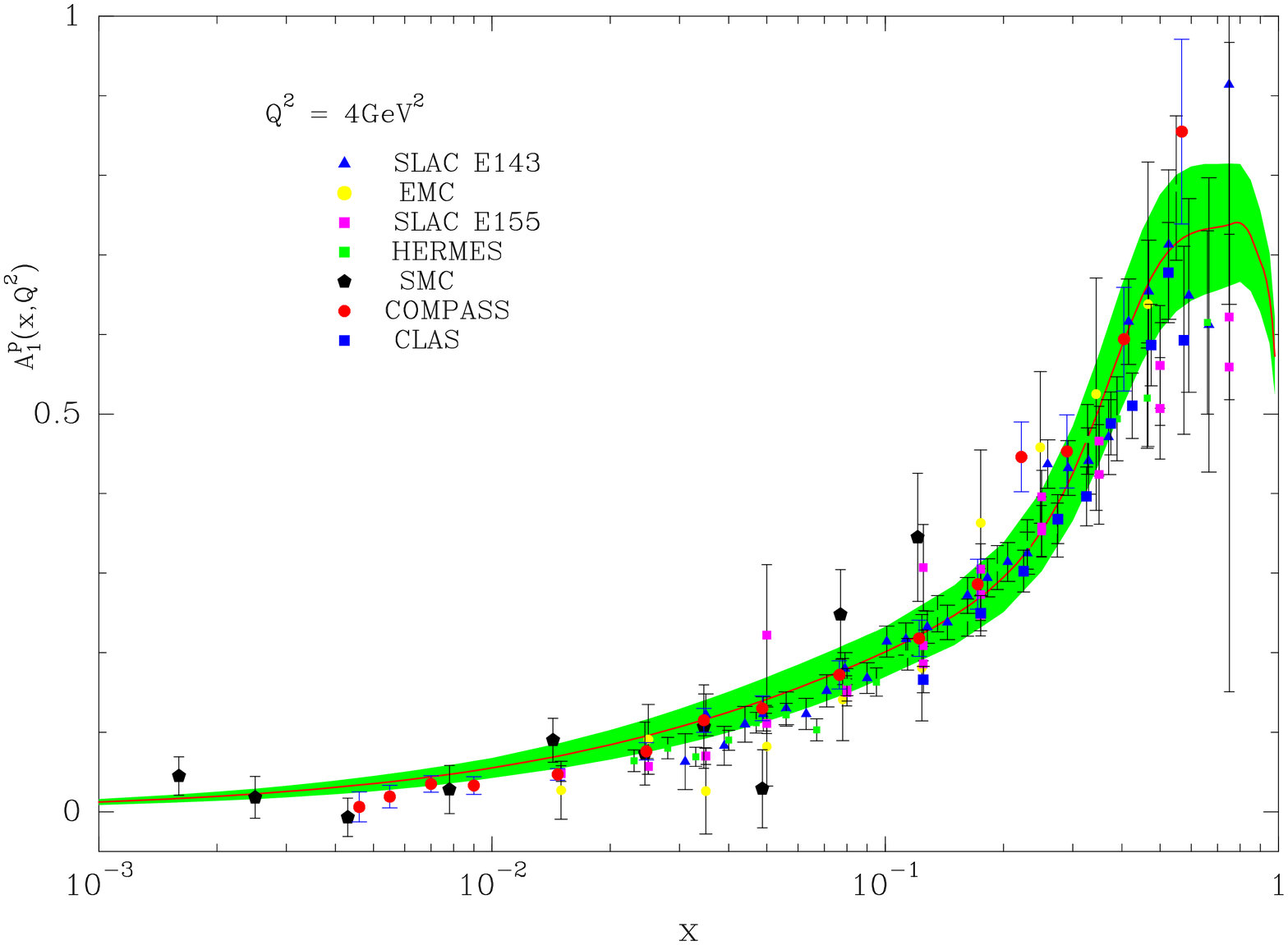}
\includegraphics[width=6.5cm]{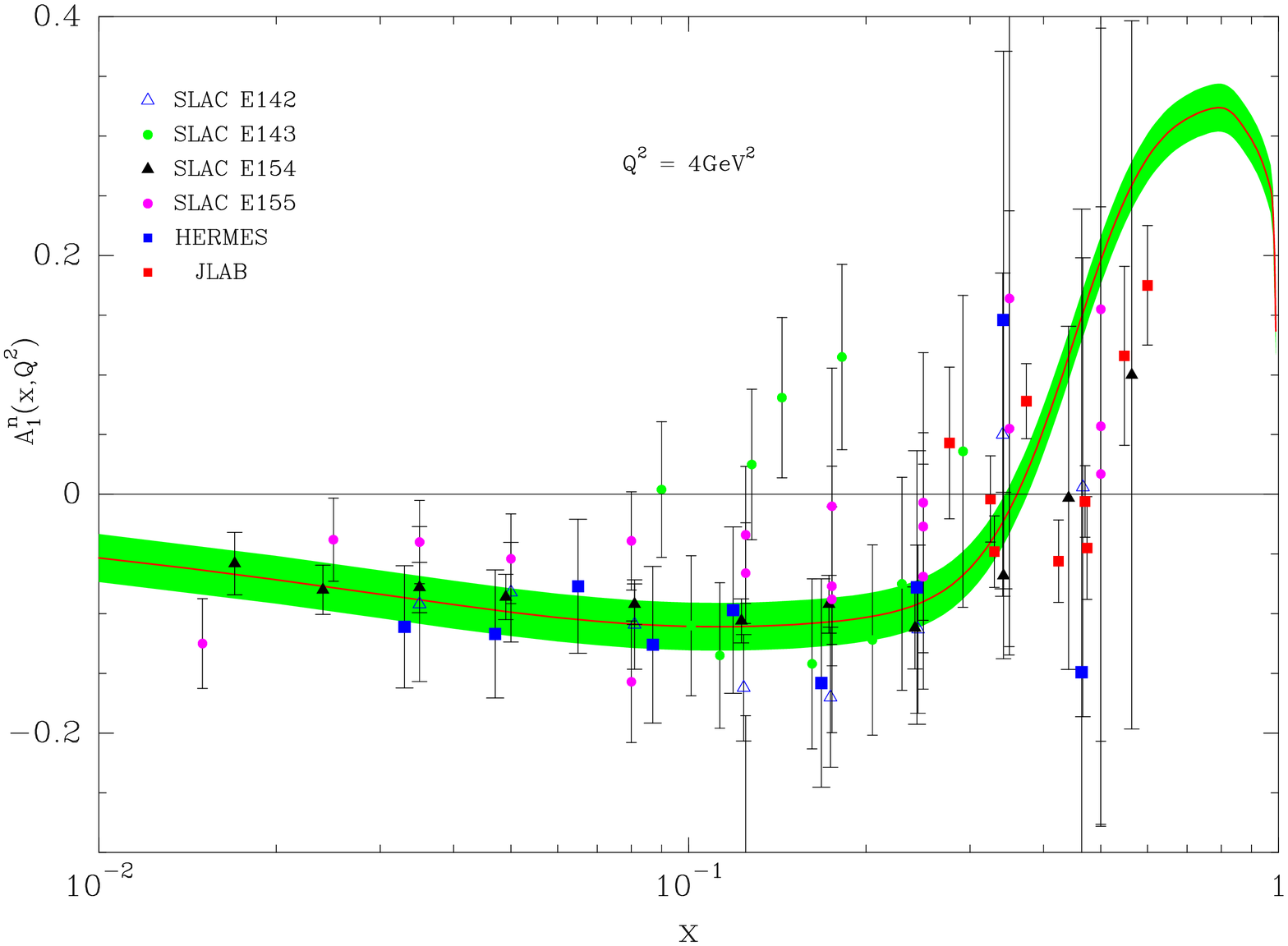}
\caption[*]{\baselineskip 1pt
 {\it Left} : Comparison of the world data on $A_1^p(x,Q^2)$ with the result of
the statistical approach at $Q^2= 4~\mbox{GeV}^2$, including the corresponding
error band. {\it Right}: Comparison of the world data on $A_1^n(x,Q^2)$ with the result
of the statistical approach at $Q^2= 4~\mbox{GeV}^2$, including the
corresponding error band.}
\label{figa1}
\end{center}
\end{figure}

We now turn to the important issue concerning
the asymmetries $A_1^{p,d,n}(x,Q^2)$, measured in polarized DIS.
We recall the definition of the asymmetry $A_1(x,Q^2)$, namely
\begin{equation}
A_1(x,Q^2)= \frac{[g_1(x,Q^2) - \gamma^2 (x,Q^2) g_2(x,Q^2)]}{F_2(x,Q^2)}
\frac{2x[1+R(x,Q^2)]}{[1+\gamma^2(x,Q^2)]}~,
\label{26}
\end{equation}
where $g_{1,2}(x,Q^2)$ are the polarized structure functions,
$\gamma^2(x,Q^2)=4x^2 M_{p}^2/Q^2$ and $R(x,Q^2)$ is the ratio between the
longitudinal and transverse photoabsorption cross sections. When $x \to 1$ for
$Q^2= 4~\mbox{GeV}^2$, $R$ is the order of 0.30 or less and $\gamma^2(x,Q^2)$
is close to 1, so if the $u$ quark dominates, we have $A_1 \sim 0.6 \Delta
u(x)/u(x)$, so it is unlikely to find $A_1 \to 1$, as required by the counting
rules prescription, which we don't impose.
We display in Fig.~\ref{figa1} the world data on $A_1^{p}(x,Q^2)$ ({\it Left})
and $A_1^{n}(x,Q^2)$ ({\it Right}), with the results of the statistical
approach at $Q^2= 4 \mbox{GeV}^2$, up to $x=1$. Indeed we find that $A_1^{p,n}
< 1$.

Finally one important outcome of this new analysis of the polarized DIS data in
the framework of the statistical approach, is the confirmation of a large positive
gluon helicity distribution, which gives a significant contribution to the
proton spin \cite{bs14}.\\

\begin{figure}[hbp]  
\begin{center}
\includegraphics[width=6.0cm]{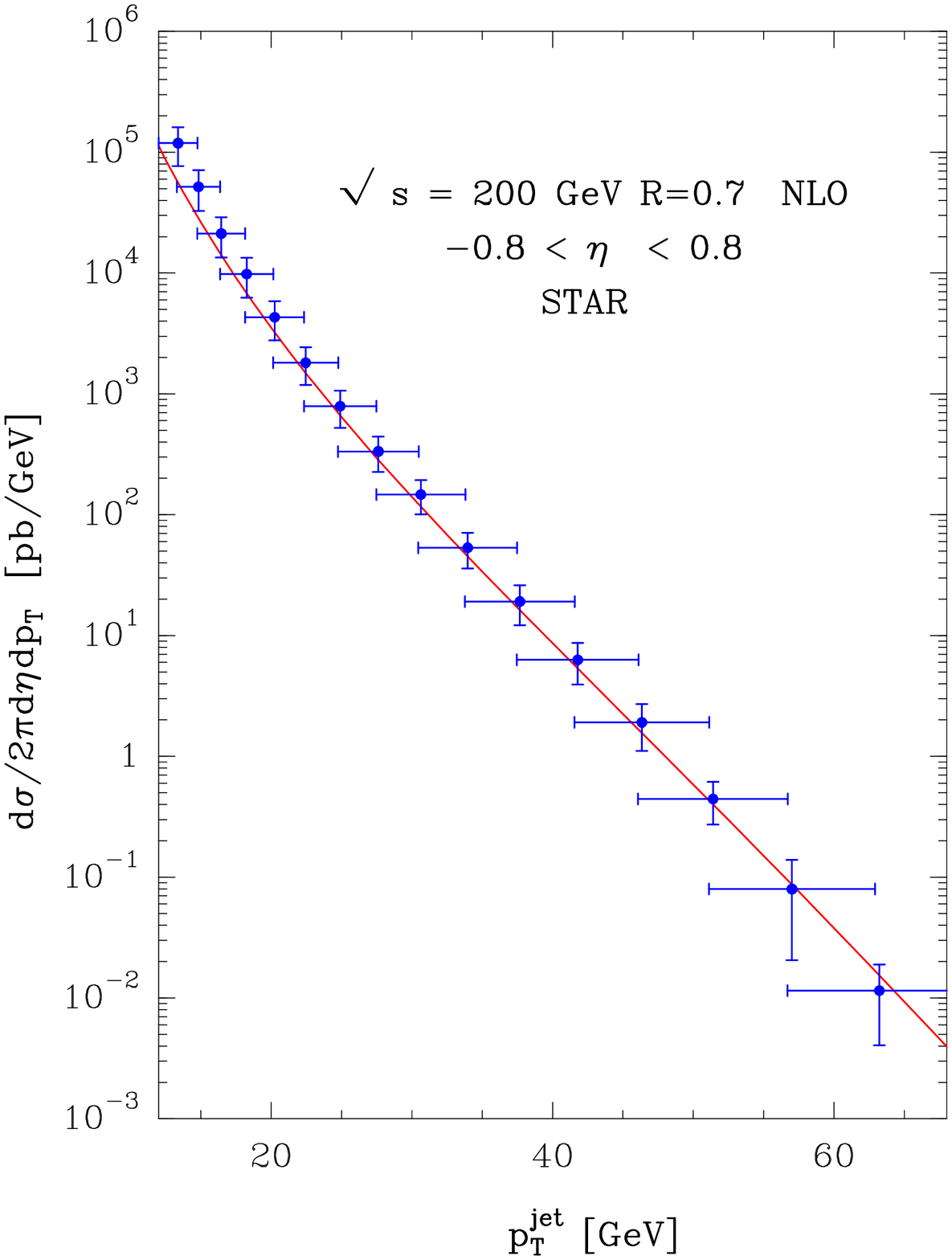}
\includegraphics[width=7.5cm]{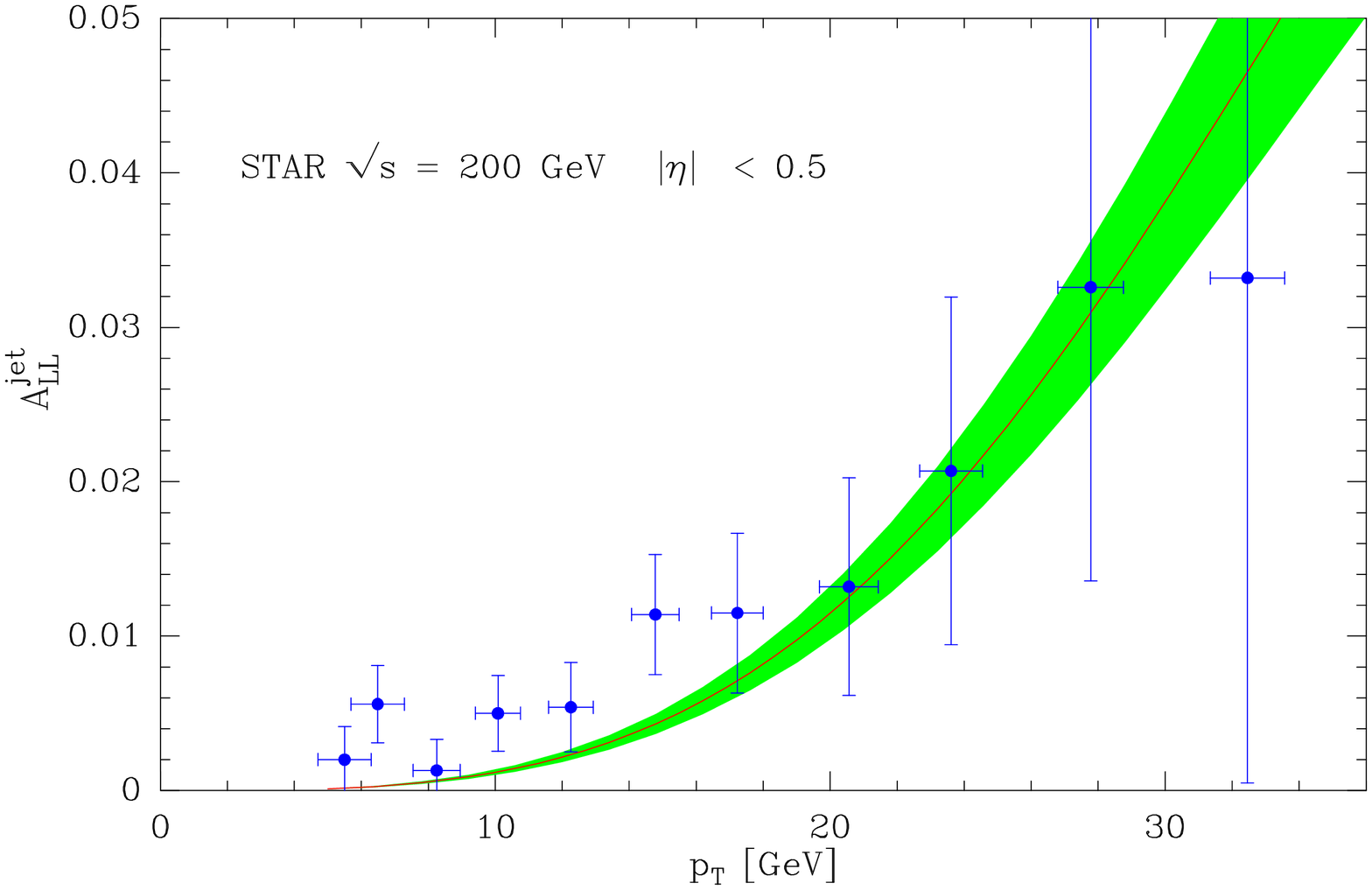}
\caption[*]{\baselineskip 1pt
 {\it Left}: Double-differential inclusive single-jet cross section in $ p p$
collisions at $\sqrt{s}$ = 200GeV, versus $p_T^{jet}$, with jet radius
parameter R=0.7, for $-0.8 <  \eta  < 0.8$, from STAR data, obtained with an
integrated luminosity of 5.39pb$^{-1}$  \cite{star1} and the prediction from
the statistical approach. {\it Right}:  Our predicted double-helicity asymmetry 
$A_{LL}^{jet}$ for single-jet production at BNL-RHIC in the near-forward rapidity 
region, versus $p_T$ and the data points from STAR \cite{star}, with the 
corresponding error band.}
\label{starjet}
\end{center}
\end{figure}

\begin{figure}[hbt]
\vspace*{-8ex}
\begin{center}
\includegraphics[width=6.2cm]{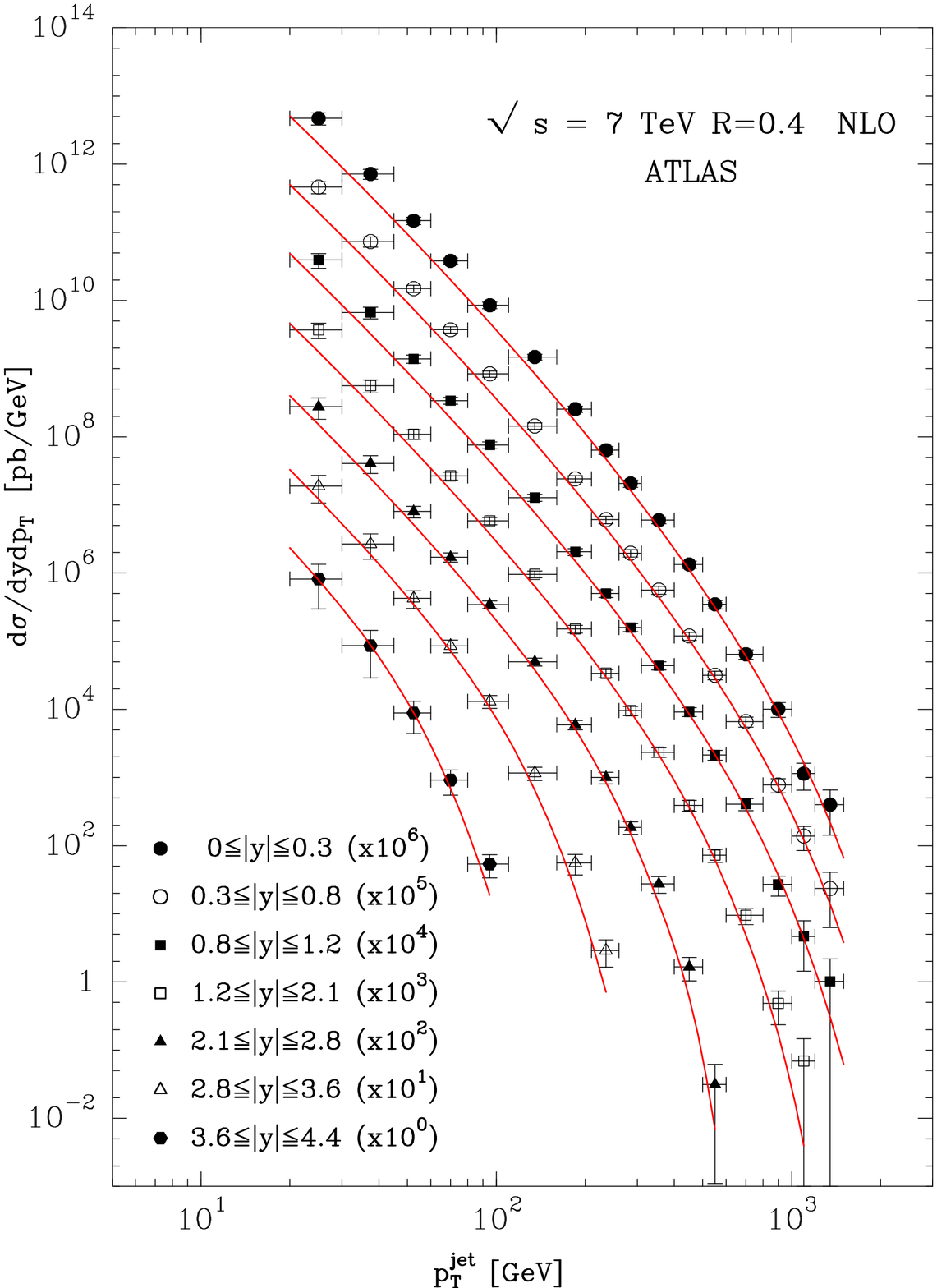}
\includegraphics[width=6.2cm]{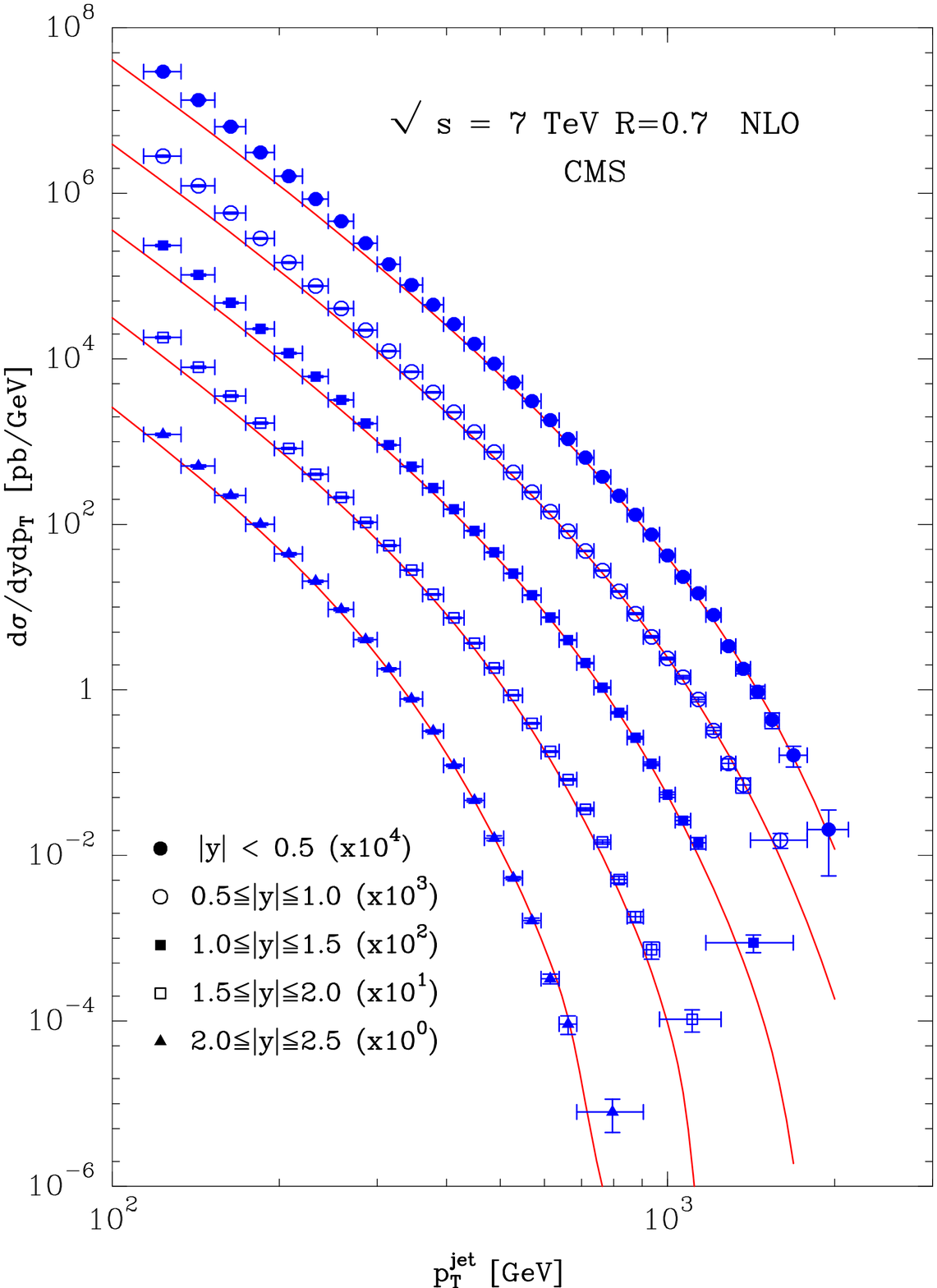}
\caption[*]{\baselineskip 1pt
 {\it Left}:  Double-differential inclusive single-jet cross section in $pp$
collisions at $\sqrt{s}$ = 7TeV, versus $p_T^{jet}$, with jet radius parameter
R = 0.4, for different rapidity bins from ATLAS \cite{atlasjet} and the
predictions from the statistical approach\\
{\it Right}: Same from CMS \cite{cms13}, with R = 0.7.}
\label{jetlhc7}
\end{center}
\vspace*{-5ex}
\end{figure}

The NLO QCD calculations at ${\cal O} ( \alpha_s^3 )$ of the cross section for the 
production of a single-jet of rapidity $y$ and 
transverse momentum $p_T$, in a $pp$ or $\bar{p}p$ collision, were done using a code
based on a semi-analytical method within the
"small-cone approximation"', improved recently with a jet algorithm for a better
definition \footnote{ We thank Werner Vogelsang for providing 
us with the code to make this calculation.}. In Fig.~\ref{starjet}({\it Left})
our results are compared with the data from STAR
experiment at BNL-RHIC and this prediction agrees very well with the 
data.\\
Now we would like to test, in a pure hadronic collision, our new positive gluon
helicity distribution, mentioned above. In a recent paper, the STAR
experiment at BNL-RHIC has reported the observation, in single-jet inclusive
production, of a non-vanishing positive double-helicity asymmetry
$A_{LL}^{jet}$ for $5 \leq p_T \leq 30$GeV, in the near-forward rapidity region
\cite{star}. We show in Fig.~\ref{starjet}({\it Right}) our prediction
compared with these high-statistics data points and the agreement is very
reasonable.\\
There are several data sets for the cross section of single-jet production, which
allow us to test our predictions, in particular the results from ATLAS and CMS displayed in
Fig. \ref{jetlhc7} at $\sqrt{s}$ = 7TeV.


{\bf Acknowledgements}\\
J. Soffer is grateful to Christophe Royon and Patrice Verdier for their kind 
invitation to EDS Blois 2015.

{}


\begin{thebibliography}{}

\bibitem{bbs1} C. Bourrely, F. Buccella and J. Soffer, Eur. Phys. J. C {\bf
23}, 487 (2002)

\bibitem{bs14} C. Bourrely and J. Soffer, Phys. Lett. B {\bf 740},  168 (2015)

\bibitem{bbs5} C. Bourrely, F. Buccella and J. Soffer, Phys. Rev. D {\bf 83},
074008 (2011)

\bibitem{bbs6} C. Bourrely, F. Buccella and J. Soffer, Int. J. of Mod. Phys. A
{\bf 28}, 1350026  (2013)

\bibitem{bbs2} C. Bourrely, F. Buccella and J. Soffer, Phys. Lett. B {\bf 648},
 39  (2007)

\bibitem{bbs4}  C. Bourrely, F. Buccella and J. Soffer, Eur. Phys. J. C {\bf
41}, 327 (2005)

\bibitem{bs15}  C. Bourrely and J. Soffer, Nucl. Phys. A {\bf 941}, 307 (2015) 

\bibitem{bbsW} C. Bourrely, F. Buccella and J. Soffer, Phys. Lett. B {\bf 726},
 296 (2013)

\bibitem{star1} B. Abelev {\it et al.} (STAR Collaboration), Phys. Rev. Lett.
{\bf 97},  252001 (2006); see also T. Sakuma, MIT Thesis (2009)

\bibitem{star} L. Adamczyk {\it et al.} (STAR Collaboration), Phys. Rev. Lett.
{\bf 115}, 9 092002 (2015)

\bibitem{atlasjet} G. Aad {\it et al.} (ATLAS Collaboration), JHEP 1502, 153 (2015); 1509, 141 (2015)

\bibitem{cms13} S. Chatrchyan $\it{et~al.}$ (CMS Collaboration), Phys. Rev. D
{\bf 87}, 112002 (2013); Erratum-ibid, 119902



\end{thebibliography}
\end{document}